Naturalistic stimuli in touch research

Anne Margarette S. Maallo*[1], Basil Duvernoy*[1], Håkan Olausson[1], Sarah McIntyre[1]

[1]Center for Social and Affective Neuroscience, Linköping University, Sweden

* AMSM and BD contributed equally

Correspondence: sarah.mcintyre@liu.se

Highlights

- Neural mechanisms of social touch are studied in reductionist laboratory conditions

- Visual neuroscience has benefitted from the use of complex naturalistic stimuli

- Naturalistic textures have also advanced discriminative somatosensory neuroscience

- Recent technical advances allow for precise design of human-to-human touch stimuli

Abstract

Neural mechanisms of touch are typically studied in laboratory settings using robotic or other types of well-controlled devices. Such stimuli are very different from highly complex naturalistic human-to-human touch interactions. The lack of scientifically useful naturalistic stimuli hampers progress, particularly in social touch research. Vision science, on the other hand, has benefitted from inventions such as virtual reality systems that have provided researchers with precision control of naturalistic stimuli. In the field of touch research, producing and manipulating stimuli is particularly challenging due to the complexity of skin mechanics. Here we review the history of touch neuroscience focusing on the contrast between strictly controlled and naturalistic stimuli, and compare the field to vision science. We discuss new methods that may overcome obstacles with precision-controlled tactile stimuli, and recent successes in naturalistic texture production. In social touch research, precise tracking and measurement of naturalistic human-to-human touch interactions offer exciting new possibilities.

Technological advances are overcoming historical barriers to tactile stimulus design, allowing the production and measurement of new kinds of naturalistic touch stimuli. Vision science has already reaped the benefits of a wide range of naturalistic stimuli and precise stimulus control. Now these benefits are becoming available to tactile researchers, and it is particularly important for the study of social and affective touch, which has been historically limited by the stimulators available.

## The problem of completely defining a tactile stimulus

Research progress in the domain of sensory systems is highly dependent on researchers' ability to produce, adapt, and manipulate stimuli in ways that are appropriate for the sensory system under investigation. In vision science, the invention of projectors, computer monitors, and virtual reality systems have provided researchers with more and more control [1] over increasingly naturalistic stimuli [2,3]. In the field of touch and somatosensation research, producing and manipulating stimuli is particularly challenging.

A reason for this challenge lies in the complexity of the skin mechanics. For any sensory system, we can think of the receptor surface as the interface between the environment and the nervous system. For vision, the receptor surface is the retina, which is limited in size, and does not typically deform. For touch, the receptor surface is the skin, which, in contrast, is distributed over the entire body, and physically deforms in response to stimulation. One way to model visual or tactile sensory stimuli is by devising a function that provides a full description of all sensory information that can potentially be seen or felt, respectively called the plenoptic and plenhaptic functions. The plenoptic function [4], defined in 1991, has a finite number of seven dimensions to capture a visual stimulus, and is used to produce computer-modelled visual environments. Because of the mechanical properties of the skin, the plenhaptic function, defined in 2011, has infinite dimensions [5], and can only be reduced to a finite number by making assumptions and simplifications of the interaction between the object and the skin.

Assuming equal levels of complexity between visual and tactile stimulators, the plenhaptic function imposes theoretical limits to the possible range of stimulation that can be mechanically achieved. This limit implies that a universal haptic stimulator is only possible with additional compromises. Thus, we might never see a single commercial product that could provide touch researchers with the power, flexibility, and fidelity of stimulus control that computer monitors and virtual reality systems provide to vision researchers. Additionally, haptic devices require physically moving parts close to or in contact with the receptor surface, which limits the spatial resolution with which researchers can manipulate the stimulus.

Fortunately, a large number of haptic interfaces for a growing number of purposes are now becoming accessible to industry and research [6–8]. Advances in computer vision and motion tracking are providing new tools for 3D-modelling of hands, bodies and objects that can be used to measure physical features of both social and non-social touch interactions [9–12]. These advances provide somatosensory researchers with new methods of systematically exploring neural processing of much more complex stimuli that bear a closer resemblance to the tactile environment humans navigate in our daily lives.

**Vision science: a benchmark for stimulus complexity**

In their seminal work, Hubel and Wiesel [1] studied receptive fields in the cat's striate cortex. Around the same time, scientists also quantified responses in the cat's somatic cortex [13]. These pioneering studies employed highly controlled and parametrized stimuli (e.g., spotlight to the retina and skin indentation for vision and somatosensation, respectively).

While the mid-20th century saw fields of somatosensory and visual research reach similar milestone discoveries, vision science has successfully pivoted to using naturalistic stimuli in recent years. The turn of the century came with improved technology that enabled researchers to explore cortical responses beyond the simplistic light patterns used previously. Advances up to this point have been discussed in detail elsewhere (for instance, see [14] and [15]). In parallel, the non-invasive whole-brain neuroimaging modality that is fMRI was slowly becoming more mainstream [16]. With the advances in stimulation and recording together, in human research in particular, we highlight below some landmark studies regarding natural stimuli in vision.

In 2003, Bartels and Zeki [2] published the first fMRI study to use uncontrolled stimuli – a James Bond movie – vis-à-vis visual function. Bartels and Zeki showed that even when stimuli are embedded in complex situations (i.e. not in isolation, as is the case in parametrized testing), functional specialization was nevertheless preserved. In another study that used free viewing of a video clip, Hasson et al. [17] highlighted the ecological validity of naturalistic stimuli in vision research. From whole-brain fMRI signals, the authors found significant inter-subject correlation of neural activity in a large swathe of the cortex, implying that brains exhibit the same spatiotemporal pattern in response to the same stimuli. Likewise, in the field of computational neuroscience, while the biologically inspired artificial neural networks (ANN) are deeply rooted in the findings of Hubel and Wiesel [18], in recent years, ANNs have moved in the direction of representational models of the natural scene [19,20].

Contemporarily, while both low-level and high-level stimuli are still used to serve different scientific objectives, more advanced technologies are becoming widely available that allow for more degrees of freedom for manipulation. For instance, consumer virtual or augmented reality (VR, AR, respectively) do not only offer visual stimulation. With VR/AR, other sensory systems can be integrated and studied in parallel.

## Stimulus design limits touch research

Historically, most psychophysical and neurophysiological studies of touch have been performed either with simple hand-held stimulators with minimal control, or with well-controlled stimuli that are very limited in the range of tactile stimulations they can produce. In the field of affective and social touch research, there is widespread use of commercially available paint brushes made with soft goat hair, which when stroked across the skin are generally felt as pleasant [21–26]. Despite being a complex stimulus composed of many individual hairs moving in slightly different ways, very little is known about the mechanical interaction between the hairs of the brush and the skin, and which physical features of the brush contribute to its softness or pleasantness. In a recent attempt to understand this by introducing mechanical changes at the skin-brush interface, the unavailability of fine stimulus control limited the precision of the conclusions [27].

Afferent neurons have been studied with precise sets of stimuli, which usually involve one fundamental physical dimension at a time (e.g.. change in just the lateral or normal force, or vibration frequency). Therefore, afferent neuron types in touch have been characterized mainly by their responses to one specific dimension embedded in simplistic stimuli rather than stimuli that are typically encountered in daily life.

Compared to neurophysiological studies, psychophysical studies have enjoyed a somewhat broader range of stimuli with better control. An example of how early technological advances in touch interfaces has improved our understanding of the somatosensory system is the development of apparent motion devices and the tactile motion aftereffect. Investigation of the visual motion aftereffect has provided insights into retinal and cortical processing of visual inputs. There are multiple ways of inducing the aftereffect with different types of visual stimuli, and these can be explained by neural activity in different locations in the nervous system [28]. Many of these insights were gleaned from the careful manipulation of visual elements on computer monitors such as random dot kinematograms, in which the proportion of randomly moving and consistently moving elements can be determined by the experimenter [29].

Early investigations of the tactile motion aftereffect suggested that it was weak, unreliable or inconsistent [30,31]. It was only after the development of apparent motion actuators, allowing finer control over individual elements in a moving stimulus, that a robust tactile motion aftereffect could be found [32]. This immediately lead to the insight that the direction-sensitive adapted neural population relies on dynamic activation in the periphery [32], and this was shortly followed by the insight that the neural population was also likely to scale their activity to the speed of motion [33].

**Advances in haptics and motion tracking can usher in a new era of touch research**

Advances in physics, chemistry, engineering and manufacturing allow new fields to grow dramatically including robotics and materials science. In the case of haptics, the increased interest in embedding haptics feedback in our daily lives (e.g., in cars, smartphones, game controllers or clothes), has inspired haptics researchers to develop new approaches to stimulating the sense of touch. One area that has grown rapidly in the last decade is research into the texture of objects and surfaces.

Texture is one of the key aspects of the world of touch. It has been a recurring challenge to render virtual tactile textures that researchers try to tackle, with a particular interest in surface displays embedded in smartphones. A major challenge for studying texture-skin interactions is that skin and most textured materials are not transparent. Opacity limits the number of sensors one can use to measure the deformation of the skin, which is fundamental information about the interaction. To overcome this challenge, researchers developed textured surfaces that are transparent, allowing imaging of the stimulus-skin interaction [34–36]. Another solution is to use non-optical sensors. Interaction between the human skin and the texture can be estimated by a frequency analysis over time using accelerometers and the force applied on the texture using force-torque sensors [37].

Even when flat surfaces have similar coefficients of friction and thermal effects, the differences in molecular structures can be felt by humans [38,39], which shows how complex it is to render high-fidelity virtual textures. In this field, many key studies investigated natural and naturalistic textures, and carefully measured their mechanical interactions with the skin [40], as well as the neurophysiological [41] and psychophysical responses [42–44]. There have been recent efforts to build databases of texture-rendering models to better understand how they can be synthesized [45], and it is now possible to virtually represent distinct granularity of textures [46].

Another issue in touch stimulation is the concept of agent interactions. Human-to-human interactions typically involve active touch by both individuals, which is not always practical in laboratory settings [47,48]. However, a recent study showed that tactile emotions can be conveyed with a robotic human-like arm [49]. It is also possible to take advantage of human-like fingertip sensors to deliver specific stimuli even though it limits the contact area to the size of a fingertip [50,51]. The advantage of this approach is to achieve a high degree of control while producing more naturalistic stimuli. For readers interested in a more thorough discussion of haptic technologies social touch, Huisman published a review dedicated to this topic [52].

Instead of mimicking human gestures with robots, another approach is to record the gestures made by a human to evaluate physical dimensions such as pressure, velocity, or contact area. Again, the problem of skin opacity arises. Unlike with surface research, it is impossible to make the skin transparent to allow image capture. However, similar approaches with non-optical sensors have proved useful. By placing accelerometers on the back of the hand, it is possible to analyze the frequency content produced when the hand engages in touch interactions [53].

Video capture of human-to-human touch can also be used to estimate contact characteristics, despite image occlusion, using motion tracking and 3D modeling of human body parts. While motion tracking technology is not new, until recently no one had been able to reliably measure touch events. Traditional motion tracking uses the 2-dimensional information from video recordings, capturing motion information about individual points identified with visual markers or landmarks [54,55], but which cannot detect collision or touch events. Recent advances in camera technology and computer vision are making it possible to detect interpersonal touch events by creating 3D models of the surfaces of the arm, fingers and hands of individuals when they engage in social touch tasks in the laboratory [9,10,48].

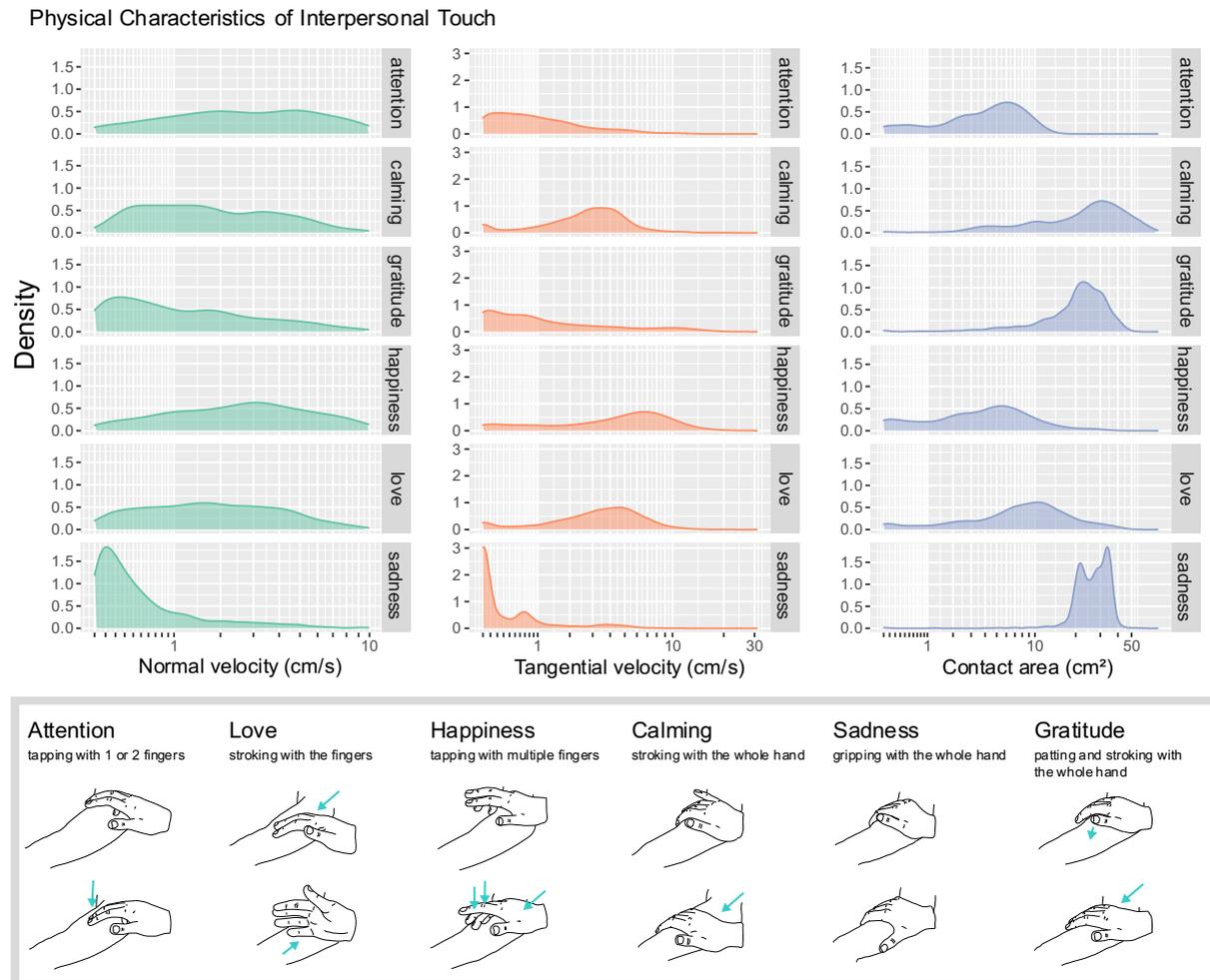

*Figure 1. Physical characteristics of real interpersonal touch, measured with motion tracking, using a 3D model of the arm, hand and fingers. These touches were performed during a touch communication task reported in [48]. Schematic representations of the touch expressions are reproduced from [48].*

With a view of precisely measuring naturalistic interactions, Figure 1 shows contact characteristics of interpersonal human-to-human touch measured with motion tracking and 3D anatomical modelling. The density plots show normal velocity (person A's hand moving down to press on person B's arm), tangential velocity (person A's hand moves along person B's arm) and contact area (the area of contact between person A's hand and fingers and person B's arm). In each of these panels, person A was trying to communicate one of six messages (attention, calming, gratitude, happiness, love, or sadness) using only touch. The density distributions reveal that the contact characteristics vary considerably, even within a single message. These results suggest socially relevant stimuli have complex, but measurable physical attributes, and that it is not appropriate to only use highly controlled and consistent stimulators if they are limited in the range of touch stimuli they can produce.

## Conclusion

Progress in the understanding of neural mechanisms of social touch is hampered by a lack of adequate stimuli. Visual neuroscience, on the other hand, has made important progress by using naturalistic stimuli. For tactile research, it is very difficult to build stimulators that can produce the highly complex skin deformation that characterizes naturalistic stimuli, although progress is being made in this area. For example, success has been achieved in the production of naturalistic surface textures, and in our subsequent understanding of texture perception and its neural basis. Another way forward for social and affective touch is to capitalize on recent improvements to 3D-modelling in motion tracking systems that can make high-precision measurement and allow identification of key-parameters in true human-to-human interactions.

## Disclosures

The authors declare no conflict of interest.

# References

1. Hubel DH, Wiesel TN: **Receptive fields and functional architecture of monkey striate cortex**. *J Physiol* 1968, **195**:215–243.

2. Bartels A, Zeki S: **Functional brain mapping during free viewing of natural scenes**. *Hum Brain Mapp* 2004, **21**:75–85.

3. Meißner M, Pfeiffer J, Pfeiffer T, Oppewal H: **Combining virtual reality and mobile eye tracking to provide a naturalistic experimental environment for shopper research**. *J Bus Res* 2019, **100**:445–458.

4. Adelson EH, Bergen JR: **The Plenoptic Function and the Elements of Early Vision**. In *Computational Models of Visual Processing*. . MIT Press; 1991:3–20.

5. Hayward V: **Is there a 'plenhaptic' function?** *Philos Trans R Soc B Biol Sci* 2011, **366**:3115–3122.

6. Duvernoy B, Topp S, Hayward V: **"HaptiComm", a Haptic Communicator Device for Deafblind Communication**. In *Haptic Interaction*. Edited by Kajimoto H, Lee D, Kim S-Y, Konyo M, Kyung K-U. Springer; 2019:112–115.

7. Flagg A, MacLean K: **Affective touch gesture recognition for a furry zoomorphic machine**. In *Proceedings of the 7th International Conference on Tangible, Embedded and Embodied Interaction*. . Association for Computing Machinery; 2013:25–32.

8. Pang G, Yang G, Pang Z: **Review of Robot Skin: A Potential Enabler for Safe Collaboration, Immersive Teleoperation, and Affective Interaction of Future Collaborative Robots**. *IEEE Trans Med Robot Bionics* 2021, **3**:681–700.

    •• A comprehensive review of collaborative robots ("cobots") covering the state of the art regarding technical and applied aspects of human-robot interaction. It discusses sensor-actuator feedback loops for dynamically adjustusting robot behavior.

9. Xu S, Xu C, McIntyre S, Olausson H, Gerling GJ: **Subtle Contact Nuances in the Delivery of Human-to-Human Touch Distinguish Emotional Sentiment**. *IEEE Trans Haptics* 2021, doi:10.1109/TOH.2021.3137833.

    • In this article, computer vision is used to model 3D motion tracked hand and arm interactions between two individuals. The technique is used to full effect, revealing socially relevant features of the measured touch interactions with much higher precision than has previously been possible.

10. Hauser SC, McIntyre S, Israr A, Olausson H, Gerling GJ: **Uncovering Human-to-Human Physical Interactions that Underlie Emotional and Affective Touch Communication**. In *2019 IEEE World Haptics Conference (WHC)*. . IEEE; 2019:407–412.

11. Lo C, Chu ST, Penney TB, Schirmer A: **3D Hand-Motion Tracking and Bottom-Up Classification Sheds Light on the Physical Properties of Gentle Stroking**. *Neuroscience* 2020, doi:10.1016/j.neuroscience.2020.09.037.

    • In this study, point markers are used to track stroking movements in 3D space, and analysis of movement dynamics show higher variability associated with social (vs. non-social) targets, and with the intention of providing pleasant touch (vs. a particular speed).

12. Xu C, Wang Y, Gerling GJ: **An elasticity-curvature illusion decouples cutaneous and proprioceptive cues in active exploration of soft objects**. *PLOS Comput Biol* 2021, **17**:e1008848.

13. Mountcastle VB, Davies PW, Berman AL: **Response properties of neurons of cat's somatic sensory cortex to peripheral stimuli**. *J Neurophysiol* 1957, **20**:374–407.

14. Reinagel P: **How do visual neurons respond in the real world?** *Curr Opin Neurobiol* 2001, **11**:437–442.

15. Felsen G, Dan Y: **A natural approach to studying vision**. *Nat Neurosci* 2005, **8**:1643–1646.

16. Muckli L: **What are we missing here? Brain imaging evidence for higher cognitive functions in primary visual cortex V1**. *Int J Imaging Syst Technol* 2010, **20**:131–139.

17. Hasson U, Nir Y, Levy I, Fuhrmann G, Malach R: **Intersubject Synchronization of Cortical Activity During Natural Vision**. *Science* 2004, **303**:1634–1640.

18. Lindsay GW: **Convolutional Neural Networks as a Model of the Visual System: Past, Present, and Future**. *J Cogn Neurosci* 2021, **33**:2017–2031.

19. Khosla M, Ngo GH, Jamison K, Kuceyeski A, Sabuncu MR: **Cortical response to naturalistic stimuli is largely predictable with deep neural networks**. *Sci Adv* 2021, **7**:eabe7547.

    • This article presents a deep neural network model for integrating audition, vision, and memory in the context of free viewing of video clips of varying durations. The study is made more attractive by the use of openly available data from the Human Connectome Project, as well as public availability of the authors' codes.

20. Peterson JC, Abbott JT, Griffiths TL: **Evaluating (and Improving) the Correspondence Between Deep Neural Networks and Human Representations**. *Cogn Sci* 2018, **42**:2648–2669.

21. Sailer U, Hausmann M, Croy I: **Pleasantness Only?: How Sensory and Affective Attributes Describe Touch Targeting C-Tactile Fibers**. *Exp Psychol* 2020, **67**:224–236.

22. Croy I, Bierling A, Sailer U, Ackerley R: **Individual variability of pleasantness ratings to stroking touch over different velocities**. *Neuroscience* 2020, doi:10.1016/j.neuroscience.2020.03.030.

•• This paper is representative of a turning point in affective touch research. The authors critically analyse the most popular and dominant experimental paradigm - using soft brushes to stroke the skin and record pleasantness ratings. They show that speed-dependence of perceived pleasantness is only reproducible at a group level, and that individuals vary considerably.

23. Löken LS, Wessberg J, Morrison I, McGlone F, Olausson H: **Coding of pleasant touch by unmyelinated afferents in humans**. *Nat Neurosci* 2009, **12**:547–548.

24. Jönsson EH, Backlund Wasling H, Wagnbeck V, Dimitriadis M, Georgiadis JR, Olausson H, Croy I: **Unmyelinated Tactile Cutaneous Nerves Signal Erotic Sensations**. *J Sex Med* 2015, **12**:1338–1345.

25. Kirsch LP, Krahé C, Blom N, Crucianelli L, Moro V, Jenkinson PM, Fotopoulou A: **Reading the mind in the touch: Neurophysiological specificity in the communication of emotions by touch**. *Neuropsychologia* 2018, **116**:136–149.

26. Novembre G, Etzi R, Morrison I: **Hedonic responses to touch are modulated by the perceived attractiveness of the caresser**. *Neuroscience* 2020, doi:10.1016/j.neuroscience.2020.10.007.

27. Rezaei M, Nagi SS, Xu C, McIntyre S, Olausson H, Gerling GJ: **Thin Films on the Skin, but not Frictional Agents, Attenuate the Percept of Pleasantness to Brushed Stimuli**. In *2021 IEEE World Haptics Conference (WHC)*. . 2021:49–54.

28. Mather G, Pavan A, Campana G, Casco C: **The motion aftereffect reloaded**. *Trends Cogn Sci* 2008, **12**:481–487.

29. Nishida S, Sato T: **Motion aftereffect with flickering test patterns reveals higher stages of motion processing**. *Vision Res* 1995, **35**:477–490.

30. Hazlewood V: **A note on failure to find a tactile motion aftereffect***. *Aust J Psychol* 1971, **23**:59–62.

31. Planetta PJ, Servos P: **The tactile motion aftereffect revisited**. *Somatosens Mot Res* 2008, **25**:93–99.

32. Watanabe J, Hayashi S, Kajimoto H, Tachi S, Nishida S: **Tactile motion aftereffects produced by appropriate presentation for mechanoreceptors**. *Exp Brain Res* 2007, **180**:577–582.

33. McIntyre S, Birznieks I, Vickery RM, Holcombe AO, Seizova-Cajic T: **The tactile motion aftereffect suggests an intensive code for speed in neurons sensitive to both speed and direction of motion**. *J Neurophysiol* 2016, **115**:1703–1712.

34. N. Huloux, L. Willemet and M. Wiertlewski: **How to Measure the Area of Real Contact of Skin on Glass**. *IEEE Transactions on Haptics* 2021, vol. 14, 2:235-241, doi: 10.1109/TOH.2021.3073747.

35. Kaneko S, Kajimoto H: **Measurement System for Finger Skin Displacement on a Textured Surface Using Index Matching**. *Appl Sci* 2020, **10**:4184.

36. Kaneko S, Kajimoto H: **Method of Observing Finger Skin Displacement on a Textured Surface Using Index Matching**. In *Haptics: Perception, Devices, Control, and Applications*. Edited by Bello F, Kajimoto H, Visell Y. Springer International Publishing; 2016:147–155.

37. Michael Wiertlewski, Vincent Hayward: **Mechanical behavior of the fingertip in the range of frequencies and displacements relevant to touch**. *Journal of Biomechanics* 2012, vol. 42 11:1869–1894.

38. Gueorguiev D, Bochereau S, Mouraux A, Hayward V, Thonnard J-L: **Touch uses frictional cues to discriminate flat materials**. *Sci Rep* 2016, **6**:25553.

39. Nolin A, Licht A, Pierson K, Lo C-Y, Kayser LV, Dhong C: **Predicting human touch sensitivity to single atom substitutions in surface monolayers for molecular control in tactile interfaces**. *Soft Matter* 2021, **17**:5050–5060.

   • These researchers made pairs of surfaces from very similar molecules that differed in just one atom, and participants could discriminate between them. The discriminable pairs were predictable based on measured frictional properties of the surfaces.

40. Manfredi LR, Saal HP, Brown KJ, Zielinski MC, Dammann JF, Polashock VS, Bensmaia SJ: **Natural scenes in tactile texture**. *J Neurophysiol* 2014, **111**:1792–1802.

41. Goodman JM, Bensmaia SJ: **A Variation Code Accounts for the Perceived Roughness of Coarsely Textured Surfaces**. *Sci Rep* 2017, **7**:46699.

42. Weber AI, Saal HP, Lieber JD, Cheng J-W, Manfredi LR, Dammann JF, Bensmaia SJ: **Spatial and temporal codes mediate the tactile perception of natural textures**. *Proc Natl Acad Sci* 2013, **110**:17107–17112.

43. Kuroki S, Sawayama M, Nishida S: *Haptic metameric textures*. Neuroscience; 2019.

44. Tymms C, Zorin D, Gardner EP: **Tactile perception of the roughness of 3D-printed textures**. *J Neurophysiol* 2018, **119**:862–876.

45. Culbertson H, López Delgado JJ, Kuchenbecker KJ: **One hundred data-driven haptic texture models and open-source methods for rendering on 3D objects**. In *2014 IEEE Haptics Symposium (HAPTICS)*. 2014:319–325.

46. Otake K, Okamoto S, Akiyama Y, Yamada Y: **Virtual tactile texture using electrostatic friction display for natural materials: The role of low and high frequency textural stimuli**. In *2021 30th IEEE International Conference on Robot Human Interactive Communication (RO-MAN)*. 2021:392–397.

47. Hauser SC, Nagi SS, McIntyre S, Israr A, Olausson H, Gerling GJ: **From Human-to-Human Touch to Peripheral Nerve Responses**. In *2019 IEEE World Haptics Conference (WHC)*. 2019:592–597.

48. McIntyre S, Hauser SC, Kusztor A, Boehme R, Moungou A, Isager PM, Homman L, Novembre G, Nagi S, Israr A, et al.: **The language of social touch is intuitive and quantifiable**. 2021, *PsyArXiv,* doi:10.31234/osf.io/smktq.

49. Teyssier M, Bailly G, Pelachaud C, Lecolinet E: **Conveying Emotions Through Device-Initiated Touch**. *IEEE Trans Affect Comput* 2020, doi:10.1109/TAFFC.2020.3008693.

50. Fishel J, Loeb G: **Bayesian Exploration for Intelligent Identification of Textures**. *Front Neurorobotics* 2012, **6**:4.

51. Sun H, Kuchenbecker KJ, Martius G: **A soft thumb-sized vision-based sensor with accurate all-round force perception**. *ArXiv211105934 Cs Eess* 2021,

52. Huisman G: **Social Touch Technology: A Survey of Haptic Technology for Social Touch**. *IEEE Trans Haptics* 2017, **10**:391–408.

53. Shao Y, Hayward V, Visell Y: **Spatial patterns of cutaneous vibration during whole-hand haptic interactions**. *Proc Natl Acad Sci* 2016, **113**:4188–4193.

54. Johansson G: **Visual perception of biological motion and a model for its analysis**. *Percept Psychophys* 1973, **14**:201–211.

55. Mathis A, Mamidanna P, Cury KM, Abe T, Murthy VN, Mathis MW, Bethge M: **DeepLabCut: markerless pose estimation of user-defined body parts with deep learning**. *Nat Neurosci* 2018, **21**:1281–1289.